\RequirePackage{fix-cm}
\documentclass[twocolumn]{svjour3}          
\smartqed  
\usepackage{amsfonts}%
\usepackage{amsmath}%
\usepackage{amssymb}%
\usepackage{graphicx}
\usepackage[usenames]{color}
\begin{document}

\title{Anderson-Bogoliubov collective excitations in superfluid Fermi gases at nonzero temperatures
}


\author{S. N. Klimin       \and
        H. Kurkjian        \and
        J. Tempere
}


\institute{S. N. Klimin         \and
           H. Kurkjian       \and
           J. Tempere  \at
           TQC, Universiteit Antwerpen, Universiteitsplein 1, B-2610 Antwerpen, Belgium \\
           \email{jacques.tempere@uantwerpen.be}
    \and
    J. Tempere \at
    Lyman Laboratory of Physics, Harvard University, Cambridge,
    Massachusetts 02138, USA
}

\date{\today}

\maketitle

\begin{abstract}
The Anderson-Bogoliubov branch of collective excitations in a condensed Fermi
gas is treated using the effective bosonic action of Gaussian pair
fluctuations. The spectra of collective excitations are treated for finite
temperature and momentum throughout the BCS-BEC crossover. The obtained
spectra explain, both qualitatively and quantitatively, recent experimental
results on Goldstone modes in atomic Fermi superfluids.

\end{abstract}

\section{Introduction}

The Anderson-Bogoliubov (AB) collective excitations (also named Goldstone
modes in the low-momentum limit) are sound-like oscillations of a superfluid
phase of condensed Bose or Fermi gases. They are already widely studied
theoretically both at $T=0$ and at nonzero temperatures. Sound modes in
superconductors were first considered by Anderson \cite{Anderson} within the
random phase approximation. Spectra of AB collective excitations for ultracold
atomic gases in the zero-temperature case have been well established within
the Gaussian pair fluctuation theory (GPF) and the RPA, both in the
long-wavelength limit \cite{Anderson,Engelbrecht,Marini,Salasnich,Diener2008}
and for nonzero phonon momentum \cite{Combescot,Kurkjian2016}, and within the
fermion-boson model \cite{Huang}. For nonzero temperatures in the
$q\rightarrow0$ limit, the sound velocity \cite{Ohashi2003,Pieri2004,Kosztin}
and damping \cite{Zhang,Zou,Kurkjian2017,Kurkjian2017-2,Klimin2011,Castin}
were predicted theoretically, and partially measured \cite{Hoinka}. The
damping of collective modes by a particle-hole continuum was recently observed
in the normal phase of Helium-3 \cite{Panholzer2012}.
Previous theoretical approches are limited to the low temperature regime,
where the spectrum can be calculated pertubatively from the zero temperature
case. The present work is focused on the energy spectrum and the damping
factor of AB modes in ultracold Fermi gases in the whole BCS-BEC crossover
range with finite momentum at nonzero temperature. The treatment is based on
the GPF effective action, which incorporates the effect of one-phonon
absorption/emission by a fermionic quasiparticle in the collective mode
spectrum. Other effects, e. g., three-and four-phonon scattering processes
\cite{Kurkjian2017} are beyond the scope of the present work. The obtained
spectra of AB collective excitations are verified by comparison with recent
experimental results \cite{Hoinka}, both in the long-wavelength limit and at a
nonzero phonon momentum. We also discuss the relation of the present approach
and obtained results to preceding works on AB collective excitations.

\section{Collective oscillation excitations in a superfluid Fermi gas}

We consider collective excitations on a superfluid Fermi gas on the basis of
the partition function which is the path integral on the bosonic pairing field
$\left(  \bar{\Psi},\Psi\right)  $%
\begin{equation}
\mathcal{Z}\propto\int\mathcal{D}\left[  \bar{\Psi},\Psi\right]
e^{-S_{\mathrm{eff}}}, \label{Z}%
\end{equation}
with the effective bosonic action $S_{\mathrm{eff}}$,%
\begin{equation}
S_{\mathrm{eff}}=-\int_{0}^{\beta}d\tau\int d\mathbf{r}\frac{1}{g}\bar{\Psi
}\left(  \mathbf{r},\tau\right)  \Psi\left(  \mathbf{r},\tau\right)
-\operatorname{Tr}\ln\left[  -\mathbb{G}^{-1}\right]  , \label{Seff1a}%
\end{equation}
where $\beta$ is inverse to temperature, and $g$ is the coupling constant for
the contact fermion-fermion interaction, which is renormalized through the
$s$-wave scattering length $a_{s}$ as in Ref. \cite{Diener2008}. Throughout
the paper, the units are $\hbar=1$, the fermion mass $m=1/2$, and the Fermi
wave vector $k_{F}=\left(  3\pi^{2}n\right)  ^{1/3}=1$, $n$ being the particle
density. The bosonic partition function for an interacting Fermi gas appears
as a result of the Hubbard-Stratonivich transformation \cite{Diener2008}. In
this effective action, $\mathbb{G}^{-1}\left(  \mathbf{r},\tau\right)  $ is
the inverse Nambu tensor,%
\begin{equation}
\mathbb{G}^{-1}\left(  \mathbf{r},\tau\right)  =\left(
\begin{array}
[c]{cc}%
-\frac{\partial}{\partial\tau}+\nabla_{\mathbf{r}}^{2}+\mu & \Psi\left(
\mathbf{r},\tau\right) \\
\bar{\Psi}\left(  \mathbf{r},\tau\right)  & -\frac{\partial}{\partial\tau
}-\nabla_{\mathbf{r}}^{2}-\mu
\end{array}
\right)  , \label{Fa}%
\end{equation}
with the chemical potential $\mu$.

The partition function is determined within this model by Eq. (\ref{Z}) with
the action (\ref{Seff1a}). Next, we consider collective excitations which are
oscillation modes of the pairing field about a uniform saddle-point value
$\Psi\left(  \mathbf{r},\tau\right)  =\Delta$:%
\begin{equation}
\Psi\left(  \mathbf{r},\tau\right)  =\Delta+\varphi\left(  \mathbf{r}%
,\tau\right)  ,\quad\bar{\Psi}\left(  \mathbf{r},\tau\right)  =\Delta
+\bar{\varphi}\left(  \mathbf{r},\tau\right)  \label{fluct}%
\end{equation}
where $\Delta$ is determined by the least action principle and satisfies the
saddle-point gap equation,%
\begin{equation}
\int\frac{d\mathbf{k}}{\left(  2\pi\right)  ^{3}}\left(  \frac{\tanh\left(
\frac{\beta E_{\mathbf{k}}}{2}\right)  }{2E_{\mathbf{k}}}-\frac{1}{2k^{2}%
}\right)  +\frac{1}{8\pi a_{s}}=0, \label{gapeq}%
\end{equation}
where $E_{\mathbf{k}}=\sqrt{\xi_{\mathbf{k}}^{2}+\Delta^{2}}$ is the
Bogoliubov excitation energy with the free-particle energy $\xi_{\mathbf{k}%
}=k^{2}-\mu$.

For the analysis of small oscillations about the least action solution, we
keep the quadratic expansion of the effective bosonic action. The quadratic
Gaussian pair fluctuation (GPF) action in the Matsubara $\left(
\mathbf{q},i\Omega_{n}\right)  $ representation is the $\left(  2\times
2\right)  $ matrix:%
\begin{align}
S^{\left(  quad\right)  }  &  =\frac{1}{2}\sum_{\mathbf{q},n}\left(
\begin{array}
[c]{cc}%
\bar{\varphi}_{\mathbf{q},n} & \varphi_{-\mathbf{q},-n}%
\end{array}
\right) \nonumber\\
&  \times\mathbb{M}\left(  \mathbf{q},i\Omega_{n}\right)  \left(
\begin{array}
[c]{c}%
\varphi_{\mathbf{q},n}\\
\bar{\varphi}_{-\mathbf{q},-n}%
\end{array}
\right)  , \label{Squad}%
\end{align}
where $\mathbb{M}\left(  \mathbf{q},i\Omega_{n}\right)  $ is the inverse
fluctuation propagator \cite{Diener2008}. The explicit expressions for matrix
elements $M_{j,k}\left(  q,i\Omega_{n}\right)  $ used in the present work can
be found in Ref. \cite{Klimin2011}.

The expansion (\ref{fluct}) with (\ref{gapeq}) is not enough to determine the
gap $\Delta$ at fixed density and scattering length. The gap and chemical
potentials in the mean-field approximation represent a joint solution of the
gap and number mean-field number equations. In general, beyond the mean-field
approximation both the gap equation and the equation of state should be
modified. However, when the temperature is not very close to $T_{c}$, the
equation of state beyond the mean-field approximation, e. g., accounting for
Gaussian fluctuations, combined with the mean-field gap equation gives a good
quantitative agreement with the Monte Carlo results, except close to the
transition temperature \cite{HLD}. Consequently, we can apply saddle-point gap
equation (\ref{gapeq}) in combination with non-mean-field equations of state
what seems to be appropriate for the experimental condition $T\approx0.5T_{c}$
of Ref. \cite{Hoinka}.

\section{Spectra of collective excitations}

Spectra of collective excitations are approximately revealed using the
spectral response function for the system described by the GPF effective
action. The spectral response function is determined in the same way as in
Ref. \cite{KurkjianPB}:%
\begin{equation}
\chi\left(  q,\omega\right)  =\frac{1}{\pi}\operatorname{Im}\frac
{M_{1,1}\left(  q,\omega+i0^{+}\right)  }{\det\mathbb{M}\left(  q,\omega
+i0^{+}\right)  }. \label{hi}%
\end{equation}
%

\begin{figure}[tbh]%
\centering
\includegraphics[
height=4.2125in,
width=3.2093in
]%
{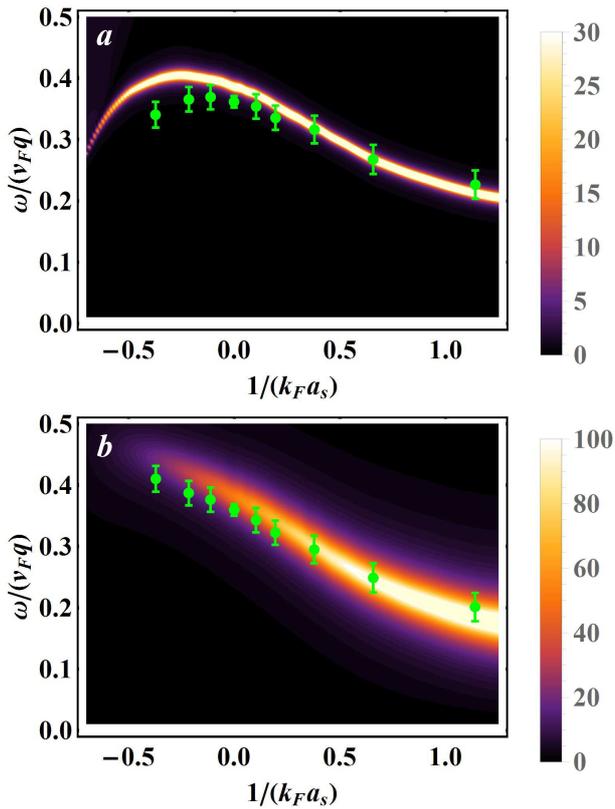}%
\caption{(Color online) Scaled spectral response function $q^{2}\chi\left(
q,\omega\right)  $ (\emph{a}) for finite momentum $q$ as indicated in
Supplement to Ref. \cite{Hoinka}, (\emph{b}) for a small $q=0.01k_{F}$. Full
dots show the experimental data of Ref. \cite{Hoinka}.}%
\label{fig:SF}%
\end{figure}

In Fig. \ref{fig:SF}, the shape of the spectral response function is compared
with two sets of the experimental results on the sound velocity for AB modes:
the raw data for nonzero $q\approx0.5k_{F}$, shown in the upper panel and the
data for a small $q\rightarrow0$, obtained in Ref. \cite{Hoinka} using a
nonzero-momentum correction and shown in the lower panel. The small-momentum
spectral response function in Fig. \ref{fig:SF}~(\emph{b}) has been calculated
for $q=0.01k_{F}$, that is sufficiently small for the comparison with the
experimental data obtained using the nonzero-momentum correction. In the limit
of small $q$, the frequency $\omega_{q}$ of the AB mode tends to the sound
wave dispersion law $\omega_{q}\rightarrow v_{s}q+O\left(  q^{3}\right)  $
with the AB mode sound velocity $v_{s}$. Consequently, we plot the spectral
response function in the variables $1/a_{s}$ and $\omega/\left(
v_{F}q\right)  $ (where $v_{F}$ is the Fermi velocity) in order to visualize
sound velocities for the comparison with the experiment.

The spectral response function has been calculated using an interpolation of
the Monte Carlo data for the zero-temperature equation of state \cite{Astr},
assuming that $\mu$ slowly varies in the range of temperatures corresponding
to the experiment $\left(  T\approx0.5T_{c}\right)  $. The gap function has
been calculated using the nonzero-temperature gap equation (\ref{gapeq}) with
that chemical potential and with the same values of the temperature as in Ref.
\cite{Hoinka}.

For a nonzero momentum, the ratio $\omega_{q}/\left(  v_{F}q\right)  $ is
smaller than $v_{s}$ at the BCS side, because the AB mode frequency is a
concave function of $q$. In the BEC case, $\omega_{q}$ is convex, and hence
$\omega_{q}/\left(  v_{F}q\right)  >v_{s}$ \cite{Kurkjian2016}. The concavity
in the BCS regime is well expressed in the figure showing a fast decrease of
the raw data of Ref. \cite{Hoinka} for the sound velocity when moving to the
BCS side in Fig. \ref{fig:SF}~(\emph{a}). Correspondingly, the same trend is
seen for the maximum of the spectral response function. The sound velocity
calculated in Ref. \cite{Hoinka} using the nonzero-momentum correction
monotonically increases when varying the inverse scattering length from BEC to
the BCS regime. However, at fixed $T$, there exists a critical value of
$1/a_{s}$ when $T=T_{c}$ [in the far BCS limit, not shown on Fig.
\ref{fig:SF}~(b)]. When approaching this value the sound velocity drops to
zero. The AB modes in this range of the inverse scattering length hardly can
be resolved experimentally due to an increasing inverse quality factor when
approaching the superfluid phase transition \cite{TBP}.

The maximum positions of the spectral response function in Fig. \ref{fig:SF}
plotted using the scaled variable $\omega_{q}/\left(  v_{F}q\right)  $ lie
rather close to the sound velocities measured in the experiment \cite{Hoinka}.
For definite conclusions, the collective excitation spectra must be determined
explicitly. To properly interpret the broadened peak of the response function
in terms of a collective excitation, one should look for the complex root of
the equation $\det\mathbb{M}\left(  q,z\right)  =0$ \cite{KurkjianPB,TBP}
where the real and imaginary part of $z$ are, respectively, the eigenfrequency
and damping factor of the AB mode. However, this equation has a priori no root
in the complex $z$ plane. To reveal a root one should perform an analytic
continuation of the function $z\rightarrow\det\mathbb{M}\left(  q,z\right)  $
through its branch cut at the real axis as proposed by Nozi\`{e}res
\cite{Nozieres} for complex poles of Green's functions. This prescription is
performed for matrix elements $M_{j,k}\left(  q,z\right)  $ of the inverse
fluctuation propagator using the spectral function, determined at the real
axis:%
\begin{equation}
\rho_{j,k}\left(  q,\omega\right)  =\lim_{\delta\rightarrow+0}\frac
{M_{j,k}\left(  q,\omega+i\delta\right)  -M_{j,k}\left(  q,\omega
-i\delta\right)  }{2i\pi}. \label{rhojk}%
\end{equation}
This spectral function is (in general, piecewise) analytic on the real axis.
It can be thus analytically extended [$\rho_{j,k}\left(  q,\omega\right)
\rightarrow\rho_{j,k}\left(  q,z\right)  $] to complex $z$ with
$\operatorname{Re}\left(  z\right)  =\omega$ and $\operatorname{Im}\left(
z\right)  <0$ from each interval where $\rho_{j,k}\left(  q,\omega\right)  $
is analytic. The analytic continuation of the matrix elements, denoted as
$M_{j,k}^{\left(  R\right)  }\left(  q,z\right)  $, is then:%
\begin{equation}
M_{j,k}^{\left(  R\right)  }\left(  q,z\right)  =\left\{
\begin{array}
[c]{cc}%
M_{j,k}\left(  q,z\right)  , & \operatorname{Im}z>0,\\
M_{j,k}\left(  q,z\right)  +2i\pi\rho_{j,k}\left(  z\right)  , &
\operatorname{Im}z<0.
\end{array}
\right.  \label{ancon}%
\end{equation}
The equation
\begin{equation}
\det\mathbb{M}^{\left(  R\right)  }\left(  q,z\right)  =0 \label{eqdet}%
\end{equation}
has complex roots in the area where $\operatorname{Im}\left(  z\right)  <0$.
These roots are denoted as $z_{q}=\omega_{q}-i\Gamma_{q}/2$, where $\omega
_{q}$ is the collective excitation frequency, and $\Gamma_{q}$ is the damping
factor. The analytic continuation method gives us frequencies and damping
factors self-consistently, i.~e. accounting for their mutual feedback, so that
the damping factor is obtained beyond the frequently used perturbation
approach (see for discussion Refs. \cite{Ohashi2003,Kurkjian2017-2,KurkjianPB}).%

\begin{figure}[tbh]%
\centering
\includegraphics[
height=4.5584in,
width=2.9144in
]%
{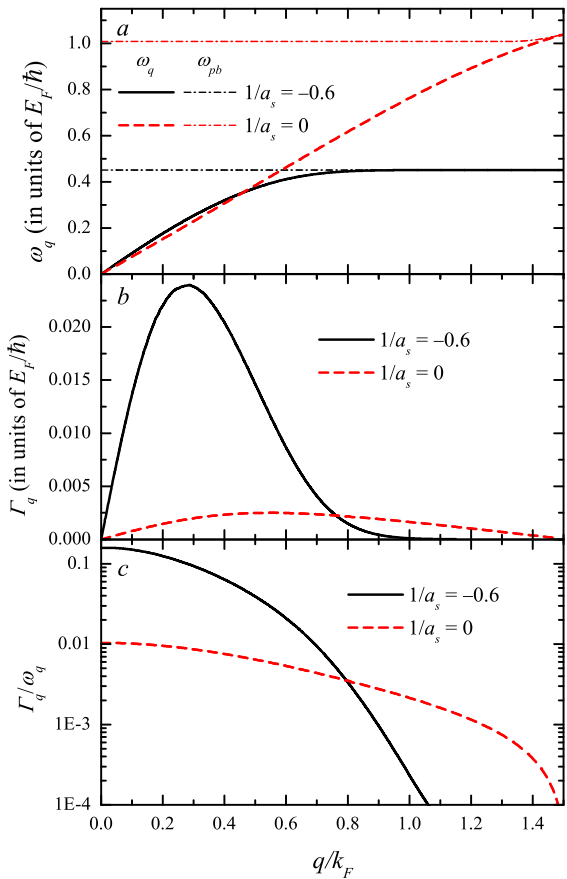}%
\caption{(Color online) (\emph{a}) Frequencies of AB modes of a superfluid
Fermi gas as a function of momentum for $1/a_{s}=-0.6$ (\emph{solid curve})
and for $1/a_{s}=0$ (\emph{dashed curve}). Dot-dashed curves: the pair
breaking threshold frequencies. (\emph{b}) The damping factor and (\emph{c})
the inverse quality factor for the same inverse scattering lengths as in the
panel (\emph{a}).}%
\label{fig:Q}%
\end{figure}

The momentum dependence of the frequencies and damping factors of the AB modes
obtained from the equation (\ref{eqdet}) is shown in Fig. \ref{fig:Q} for two
cases relevant for the experiment \cite{Hoinka}: in the BCS regime with
$1/a_{s}=-0.6$ and at unitarity, $1/a_{s}=0$. Like above, the background
parameters $\left(  \mu,\Delta\right)  $ are found from the Monte Carlo
equation of state and the finite-temperature gap equation (\ref{gapeq}). The
momentum dependence of the AB mode frequency is qualitatively the same as in
preceding works \cite{Kurkjian2016,Ohashi2003,Pieri2004}, but quantitatively
differs from them because we use different background parameters, and we
include the nonzero temperature energy shift. The AB mode energy tends to the
pair-breaking threshold energy for considered values of momentum when
increasing $q$. The damping factor exhibits a maximum at nonzero $q$ and
diminishes when the excitation energy approaches the pair-breaking threshold.
The momentum dependence of the inverse quality factor $\Gamma_{q}/\omega_{q}$
for AB modes is similar to that obtained in Ref. \cite{Kurkjian2017-2} (where
the AB mode spectra were determined using mean-field background parameters and
within a perturbative approximation).

When approaching the pair-breaking continuum, the terms of $M_{j,k}\left(
q,z\right)  $ which describe the breaking of a pair into fermionic
quasiparticles (terms with denominators $z\pm\left(  E_{\mathbf{k}%
}+E_{\mathbf{k}+\mathbf{q}}\right)  $, see \cite{Kurkjian2017-2}) become
almost resonant and repel the AB branch, forbidding it to enter the continuum.
Since the branch stays outside the continuum, these terms are never exactly
resonant and so, never contribute to the damping rate. Still, they render the
terms describing absorption-emission processes [with denominators $z\pm\left(
E_{\mathbf{k}}-E_{\mathbf{k}+\mathbf{q}}\right)  $] negligible, what explains
the suppression of the absorption-emission damping rate on Fig. \ref{fig:Q}
when approaching the pair-breaking continuum.%

\begin{figure}[ptbh]%
\centering
\includegraphics[
height=3.9557in,
width=2.8816in
]%
{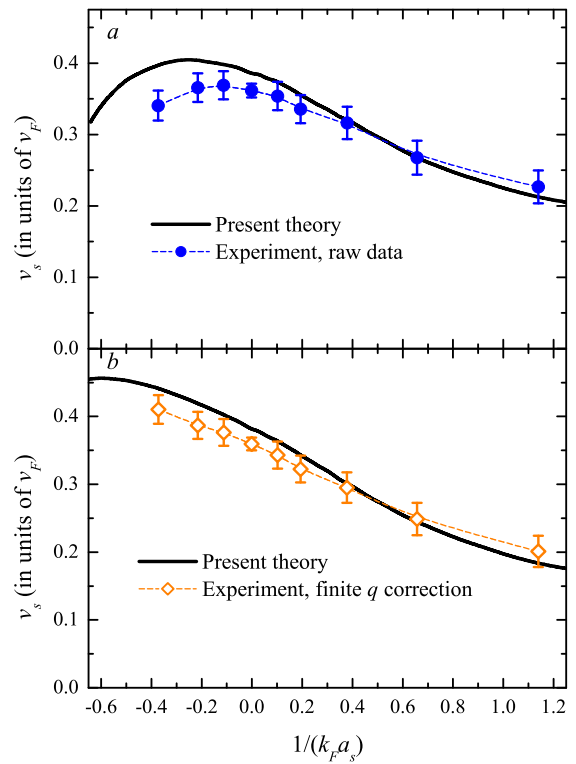}%
\caption{(Color online) (\emph{a}) Scaled AB mode frequencies $\omega
_{q}/\left(  v_{F}q\right)  $ obtained from the equation for the collective
excitations with finite $q$ determined in Ref. \cite{Hoinka}. The symbols show
the experimental results of Ref. \cite{Hoinka}. (\emph{b}) The same with
$q=0.01k_{F}$ used in our calculation, compared with the finite $q$ correction
result of the experiment.}%
\label{fig:Comp}%
\end{figure}

The comparison of sound velocities obtained using the analytic continuation of
the inverse fluctuation propagator with the experimental data of Ref.
\cite{Hoinka} is shown in Fig. \ref{fig:Comp}. As can be seen from Fig.
\ref{fig:Comp}, the calculated sound velocities are in good agreement with the
experiment \cite{Hoinka}. A difference between the calculated and measured
sound velocities can be attributed to several reasons: an inaccuracy of the
experimental determination of input parameters (e. g., the temperature and the
momentum), a difference of the chemical potential from its precise
nonzero-temperature values, and possibly an influence of induced interactions,
which can be significant in the BCS regime \cite{GMB}.

\section{Conclusions}

In the present work, we analyze spectra and damping factors for
nonzero-momentum AB collective excitations in superfluid Fermi gases as a
function of the temperature, momentum and the interaction strength. The
treatment is based on the effective Gaussian pair fluctuation action for the
pairing field. This approach is well substantiated when anharmonicity effects
can be neglected.

The energy spectrum of AB modes has been qualitatively shown using the pairing
field spectral response function. Then, quantitative results for the AB mode
spectra have been obtained from complex roots of the analytically continued
determinant of the inverse fluctuation propagator. This method provides a
self-consistent non-perturbative solution for the AB mode frequency and the
damping factor.

The experimental sound velocity is compared in Ref. \cite{Hoinka} with several
theoretical predictions \cite{Diener2008,Manini,Haussmann,Tajima}. In order to
clarify the novelty of the present work, it is worth noting that they concern
the sound velocity obtained in the precise $q\rightarrow0$ limit at $T=0$,
while the present work is focused at the $q\neq0$ behavior of AB modes at a
nonzero temperature, what is more appropriate for a comparison with the
experiment. The existing theory of AB modes for $q\neq0$, e.~g.,
\cite{Combescot,Kurkjian2016} exploits the mean-field equation of state, what
also favors the relevance of the present study, where more realistic equations
of state are used.

The AB mode spectra have been calculated using reliable background parameters
obtained accounting for fluctuations. As a result, calculated sound velocities
of AB modes exhibit a good agreement with experimental data.

\begin{acknowledgements}
The present work is supported by the University Research Fund (BOF) of the
University of Antwerp and by the Flemish Research Foundation (FWO-Vl), project
No. G.0429.15.N and the European Union's Horizon 2020 research and innovation
program under the Marie Sk\l odowska-Curie grant agreement No. 665501.
\end{acknowledgements}

\end{document}